\documentclass[preprint,letterpaper,tightenlines,nofootinbib]{revtex4}

\usepackage{amsmath}
\usepackage[bookmarks=false]{hyperref}

\newcommand{\eq}[1]{Eq.~\eqref{eq:#1}}
\newcommand{\eqs}[2]{Eqs.~\eqref{eq:#1} and \eqref{eq:#2}}
\newcommand{\eqss}[3]{Eqs.~\eqref{eq:#1}, \eqref{eq:#2}, and \eqref{eq:#3}}
\renewcommand{\sec}[1]{Sec.~\ref{sec:#1}}

\newcommand{\subsec}[1]{Sec.~\ref{subsec:#1}}

\newcommand{\ie}{i.e.}
\newcommand{\eg}{e.g.}
\newcommand{\cf}{cf.}

\newcommand{\abs}[1]{\lvert#1\rvert}
\newcommand{\mae}[3]{\langle#1\lvert#2\rvert#3\rangle}
\newcommand{\Mae}[3]{\bigl\langle#1\bigl\lvert#2\bigr\rvert#3\bigr\rangle}
\newcommand{\bra}[1]{\langle#1\rvert}
\newcommand{\ket}[1]{\lvert#1\rangle}

\renewcommand{\vec}[1]{\mathbf{#1}}

\newcommand{\ti}[1]{ \tilde{#1}}

\newcommand{\dslash}[1]{ #1 \mspace{-9mu} /}
\newcommand{\dfslash}{\dslash{\df}}
\newcommand{\deltaslash}{ (2 \pi)^4\delta}

\newcommand{\arctanh}{{{\rm tanh}^{-1}}\!}

\newcommand{\img}{\mathrm{i}}
\newcommand{\df}{\mathrm{d}}

\newcommand{\cO}{ \mathcal{O} }

\newcommand{\cQ}{ \mathcal{Q} }

\newcommand{\mcdot}{\!\cdot\!}

\newcommand{\nslash}{ \dslash{n} }
\newcommand{\bnslash}{ \dslash{\bar{n}} }

\newcommand{\nn}{\nonumber}

\newcommand{\Ecm}{E_\mathrm{cm}}

\newcommand{\Tr}{\mathrm{Tr}}

\newcommand{\jetset}{{\{\mathrm{jets}\}}}

\newcommand{\out}{\mathrm{out}}
\newcommand{\ins}{\mathrm{in}}
\newcommand{\incl}{\mathrm{incl}}
\newcommand{\meas}{\mathrm{meas}}
\newcommand{\cusp}{\!\mathrm{cusp}}
\newcommand{\ijpair}{\langle i,j \rangle}
\newcommand{\pair}[2]{\langle #1,#2 \rangle}
\newcommand{\jetR}{{\Delta \mathcal R}}

\newcommand{\T}{\mathrm{T}} 
\newcommand{\vecT}{{\vec T}} 

\newcommand{\CA}{C_A}
\newcommand{\NF}{N_f}
\newcommand{\CF}{ C_F}

\newcommand{\as}{\alpha_s}

\newcommand{\ts}{\thinspace}
\newcommand{\dlog}[1]{\frac{\df#1}{\df\ts\ln\mu}}
\renewcommand{\ln}{\text{ln}\ts}
\newcommand{\new}{\nonumber\\}
\newcommand{\partset}{\{\mathrm{partons}\}}
\newcommand{\softfn}{S(k_{\out}^0,\{n_i, k_i^+ \}; \mu)}
\newcommand{\plus}{\!\!\!+}
\newcommand{\kout}{k_{\out}^0}

\begin{document}


\title{Factorization of Boosted Multijet Processes for Threshold Resummation}

\author{Christian W. Bauer}

\author{Nicholas Daniel Dunn}

\author{Andrew Hornig}

\affiliation{Ernest Orlando Lawrence Berkeley National Laboratory,
University of California, Berkeley, California 94720, USA
\vspace{2ex}}

\begin{abstract}

Explicit applications of factorization theorems for processes at hadron colliders near the hadronic endpoint have largely focused on simple final states with either no jets (\eg, Drell-Yan) or one inclusive jet (\eg, deep inelastic scattering and prompt photon production). Factorization for the former type of process gives rise to a soft function that depends on timelike momenta, whereas the soft function for the latter type depends on null momenta.
We derive in soft-collinear effective theory a factorization theorem that allows for an arbitrary number of jets, where the jets are defined with respect to a jet algorithm, together with any number of non-strongly interacting particles. We find the soft function in general depends on the null components of the soft momenta inside the jets and on a timelike component of the soft momentum outside of the jets. This generalizes and interpolates between the soft functions for the cases of no jets and one inclusive jet. We verify consistency of our factorization theorem to $\cO(\as)$ for any number of jets. While in this paper we demonstrate consistency only near the hadronic endpoint, we keep the kinematics general enough (in particular allowing for nonzero boost) to allow for an extension to partonic threshold resummation away from the hadronic endpoint.  

\end{abstract}

\maketitle


\section{Introduction}
\label{sec:intro}

Factorization of cross sections is the basis of every theoretical prediction at hadron colliders. In its simplest form, factorization states the measured hadronic cross section $\sigma$ can be obtained by convolving a perturbatively calculable cross section $\hat \sigma$ with nonperturbative parton distribution functions (PDFs) \cite{Collins:1981ta, Collins:1981uk},
\begin{equation}
\label{eq:ffsigma}
\sigma = f \otimes f \otimes \hat \sigma
\,.\end{equation}
The PDFs are universal, and can therefore be extracted from one process and used to make predictions in another. Moreover, $\hat \sigma$ will in general depend on a hard scale $Q$ (for example, the partonic center-of-mass energy $\sqrt{\hat{s}}$), while the PDFs depend on the scale at which they are measured, say $Q_0$. The evolution of the PDFs between these two scales resums logarithms of  $Q/Q_0$. This basic paradigm illustrates the two main uses of factorization: separation of universal, nonperturbative contributions to a cross section from perturbatively calculable contributions, and resummation of logarithms of ratios of scales to which each contribution is sensitive.

When $\hat \sigma$ depends only on a single scale, \eq{ffsigma} is the end of the story. The situation is more involved when $\hat \sigma$ itself depends on multiple, widely disparate scales. For example, in many collider physics processes involving jets, $\hat \sigma$ can depend on mass scales associated with the jets such as $M_J$, hard scales like $\sqrt{\hat{s}}$, and seesaw scales like $M_J^2/\sqrt{\hat{s}}$. In such cases large logarithms of ratios of these scales can spoil the convergence of the fixed order perturbative expansion of $\hat \sigma$. One must further factorize $\hat \sigma$ in order to resum these large logarithms and, perhaps, to separate out any other nonperturbative physics that is not captured in the PDFs~\cite{Collins:1987pm, Collins:1989gx, Kidonakis:1997gm,Kidonakis:1998bk}. The general structure of resummation, at the level of next-to-leading logarithms, has been previously investigated in \cite{Bonciani:2003nt}.

In this paper we will focus on so-called threshold logarithms. When a process approaches its kinematical threshold, there is limited phase space available for radiation. This gives rise to an incomplete cancellation between real and virtual diagrams, resulting in large logarithmic terms. This is common in situations where the invariant mass of the final state is near the maximum available energy, which limits the amount of energy that can go into excess radiation. Examples of this type of resummation can be found for Drell-Yan, deep-inelastic scattering (DIS), $B$ meson decay, and top production \cite{Altarelli:1982kh, Sterman:1986aj, Sterman:1986fn, Catani:1989ne, Catani:1990rp, Contopanagos:1993xh, Bigi:1993ex, Falk:1993vb, Korchemsky:1994jb, Laenen:1993xr, Akhoury:1995fp, Berger:1995xz, Kidonakis:1998bk, Leibovich:1999xfa, Bauer:2000ew, Manohar:2003vb, Becher:2006mr, Chen:2006vd}. It has been suggested \cite{Appell:1988ie,Catani:1998tm} that a similar effect occurs at hadron colliders away from hadronic endpoint due to the steepness of parton luminosities, and this effect was explored more quantitatively in \cite{Becher:2007ty}. In this paper, we will concentrate on hadronic threshold and assume that the invariant mass of the final state is near the maximum allowed by the collider; however, we plan on exploring resummation away from the hadronic endpoint in future work \cite{threshold2}. For this reason, we will derive a factorization theorem that can be applied away from hadronic threshold without loss of information.

An extremely useful tool to prove factorization is effective field theory. 
In the case of jet physics, soft-collinear effective theory (SCET) \cite{Bauer:2000ew, Bauer:2000yr, Bauer:2001ct, Bauer:2001yt} is the relevant effective field theory that can be used to derive factorization in  many hard scattering processes \cite{Bauer:2002nz}. The SCET Lagrangian is constructed by integrating out all modes of QCD except for soft modes and collinear modes with respect to some fixed number of directions $n_i$. 
Matching QCD onto SCET gives rise to a hard function that contains the physics of the hard scales in the problem, and matrix elements of the remaining soft and collinear fields give rise to soft and jet functions, respectively.

The first applications of SCET involved cases with particularly simple jet definitions, such as in Drell-Yan \cite{Idilbi:2005ky,Becher:2007ty} where there are no jets, hemisphere jets in event shapes \cite{Bauer:2002ie,Bauer:2003di,Lee:2006fn,Lee:2006nr,Fleming:2007qr,Schwartz:2007ib,Fleming:2007xt,Bauer:2008dt,Becher:2008cf,Hornig:2009kv,Hornig:2009vb}, or completely inclusive jets as in $B \to X_s \gamma$ \cite{Bauer:2000ew}, DIS \cite{Manohar:2003vb,Becher:2006mr,Chen:2006vd}, and prompt photon production \cite{Becher:2009th}. Factorization of jets defined with more generic algorithms was considered in \cite{Bauer:2008jx} and two-jet rates defined with jet algorithms were computed using SCET in \cite{Bauer:2003di,Trott:2006bk,Cheung:2009sg}. A study of various different jet algorithms and the dependence on the jet parameters in the framework of SCET was discussed in~\cite{Cheung:2009sg}. More recently, an NLL analysis of jet shapes in multijet events using modern jet algorithms in $e^+e^-$ collisions was performed in  \cite{Ellis:2009wj,Ellis:2010rw}. 

The goal of this paper is to derive a factorization formula for an arbitrary number of jets in the presence of any number of non-strongly interacting particles in the threshold limit.
We allow for a nonzero total rapidity and calculate the ingredients of this formula to allow resummation of threshold logarithms at NLL accuracy. This is conceptually distinct from the case of a single final state jet, which can be measured indirectly by simply demanding that a non-strongly interacting particle is produced with nonzero $p_\T$. When there is more than one final state jet, jet algorithms must be used to identify jets, and so the technology of incorporating jet algorithms into a factorization formula, developed in \cite{Bauer:2008jx} and applied to $e^+e^-$ collisions in \cite{Ellis:2010rw,Ellis:2009wj}, must be employed. The consistency of this factorization (that is, the fact that the cross section is independent of the factorization scale $\mu$) is only demonstrated here in the true hadronic endpoint. However, we plan on investigating the consistency of this factorization away from hadronic endpoint using the steepness of parton luminosities in \cite{threshold2}.

In this paper, we will assume that Glauber modes do not contribute at leading order in the power counting. This cancellation has only been formally proven at the level of cross sections for simple processes, \eg \ts\cite{Collins:1983ju}. While Glauber modes could potentially spoil our factorization theorem, we assume that for sufficiently inclusive observables the argument in \cite{Collins:1983ju} generalizes. We also assume that PDFs can always be factorized from the partonic cross section, \ie\ts that
\begin{equation}
\label{eq:ffsigma}
\frac{\df\sigma}{\df\mathcal{O}} = f \otimes f \otimes \frac{\df\hat \sigma}{\df\mathcal{O}}
\,.\end{equation}
is always valid. This has also not been proven to be generically true, but is phenomenologically valid for a wide range of observables.

Factorization formulas for the case of a single, inclusive jet have soft functions that depend on the null component of the total soft momentum in the direction of the jet. When there are no jets (\eg, Drell-Yan), the soft function depends on the timelike component of the total soft momentum.
In extending threshold resummation to more than one jet using jet algorithms, we find a soft function that depends on the timelike component of the total soft momentum outside of the jets and on the null component of the soft momentum within each of the jets. Thus, our result reduces to the previously considered cases of zero and one inclusive jet when our jet algorithm either includes none or all of the final state soft momentum, respectively.

The organization of this paper is as follows. In \sec{kinematics}, we define precisely what we mean by threshold production of $N$ jets and discuss the corresponding kinematics. In \sec{nonglobal}, we discuss how the definition of threshold affects the logarithmic structure of the result and how, in particular, our definition should not introduce so-called non-global logarithms. In \sec{algorithm}, we briefly discuss different classes of jet algorithms used at hadron colliders. We then derive our factorization theorem in \sec{factorization}, beginning for notational simplicity with the case of a single (quark or gluon) final state jet, then extending these results to the case of $N$ jets. We derive the anomalous dimensions for the objects that appear in our $N$-jet factorization formula in \sec{NLOresults} and use these in \sec{consistency} to show that our factorization theorem is formally consistent, at least in the hadronic endpoint region. Finally, we present our conclusions in \sec{conclusion}.

\section{Kinematics of Threshold Resummation}
\label{sec:kinematics}

To explain our approach to threshold resummation, how it includes both the cases of Drell-Yan and direct gauge boson production as limiting cases, and how it is extendable to arbitrary $N$-jet production, we first discuss the kinematics. By demanding that the final state contains $N$ jets each with fixed  transverse momentum $p_\T$  and pseudorapidity $\eta$, together with some number of non-strongly interacting particles with total 4-momentum $q$, we are requiring that there is a minimum partonic center-of-mass energy
\begin{equation}
\label{eq:Mmin}
\hat s_{\rm min} = \Big(q+ \sum_i^N p_{J}^i \Big)^2\,,
\end{equation}
where $p_J^i$ is the momentum of the $i$th jet. This momentum is defined in terms of the $p_J^\T$ and $\eta_J$ of the jet as
\begin{equation}
\label{eq:masslessPJ}
p_J \equiv (p_J^\T \cosh \eta_J, \vec{p}_J^\T, p_J^\T \sinh \eta_J)
\,.\end{equation}
Of course, the actual partonic center-of-mass energy $\hat s$ typically exceeds this minimum value, and in general can be as large as the available machine center-of-mass energy $s$. Therefore, the dimensionless variable $z$, defined as
\begin{equation}
\label{eq:zdefn}
z \equiv \frac{\hat s_{\rm min}}{\hat s}
\,,\end{equation}
can range from 
\begin{equation}
\tau \le z \le 1\,, \qquad {\rm with} \qquad \tau \equiv \frac{\hat s_{\rm min}}{s}
\,.\end{equation}

Going to hadronic threshold ($\tau \to 1$) 
forces $z \to 1$, such that the only emissions kinematically allowed are collinear radiation off the hard partons that form jets, as well as soft radiation. Radiation collinear to one of the jets with momentum scaling as $\Ecm (1, \lambda^2,  \lambda) $ (in the light-cone coordinates of the jet) and soft radiation scaling as $ \Ecm( \lambda^2, \lambda^2, \lambda^2)$, each contribute an equal amount to $\hat s$, where $\lambda\sim\sqrt{1-z}$ is a small, dimensionless parameter.
In this limit of restricted radiation, partonic momentum conservation can be written as
\begin{equation}
\label{eq:p-cons}
p_I  = q + k_s + \sum_i^N p_c^i
\,,\end{equation}
where $p_I $ is the total initial-state (partonic) momentum, $k_s$ is the total soft momentum, and $p_c^i$ is the momentum carried by collinear fields in the direction of jet $i$. The total momentum can be separated into two components: the first is the minimum momentum needed to create $N$ jets of fixed $p_\T$ and $\eta$ together with the non-strongly interacting particles of total momentum $q$, while the second brings the invariant mass of the final state above its minimum value $\hat s_{\rm min}$. To do this, we note that an arbitrary four-vector $p$ can be written as the sum of a massless four-vector which characterizes the  transverse momentum and pseudorapidity of $p$ and a purely timelike four-vector with a magnitude equal to the $+$-component of $p$ in light-cone coordinates about $n = (1, \vec p/\abs{\vec{p}})$  (\ie, $p^+ \equiv p^0 - \abs{\vec{p}}$), which characterizes the off-shellness of $p$. That is, for any four-vector $p$, we can write
\begin{equation}
\label{eq:p-decomp}
p^\mu = p_J^\mu + p^+ v^\mu 
\,,\end{equation}
 where $v^\mu = (1, \vec 0)$ and $p_J$ is given in \eq{masslessPJ} with $p_J^\T $ and $\eta_J$ the transverse momentum and pseudorapidity of $p$, respectively. We want to apply this relation to the total 4-momentum in each of the jets. To do this, we note that the jet algorithm will group some of the soft momentum $k_s$ into parts that belong to jet $i$, $k_i$, and a part that is not included in any of the jets, $k_\out$,
 \begin{equation}
 k_s = \sum_i^N k_i + k_\out
\,. \end{equation}
 Using this together with the relation \eq{p-decomp} and the fact that 
 \begin{equation}
 (p_c^i + k_i)^+ \equiv p_c^{i, 0} + k_i^0 - \abs{\vec{p}_c^i + \vec{k}_i} =  p_c^{i +} + k_i^+ + \cO(\lambda^4)
 \,,\end{equation} 
 where on the right-hand side, $k_i^+$ is plus with respect to $p_J$ and $p_c^{i+}$ is plus with respect to $p_c$, we can write momentum conservation \eq{p-cons} at leading order in $\lambda$ as
 \begin{equation}
\label{eq:p-cons2}
p^\mu_I  = q^\mu + k^\mu_\out + \sum_i^N p^\mu_J + v^\mu \Big[ \sum_i^N (p_i^+ + k_i^+) \Big]
\,.\end{equation}
Here, we have also used that out-of-jet collinear radiation is power suppressed \cite{Ellis:2010rw}. 

Given these definitions, we can write
\begin{align}
\label{eq:1-z}
1- z &= \frac{2}{\hat s} p_I \cdot \bigg( k_\out +  v \Big[ \sum_i^N (p_i^+ + k_i^+) \Big] \bigg)+ \cO(\lambda^4) \nn\\
& = \frac{2}{\hat s} \Big(p_I \cdot k_\out   +  p_I^0 \sum_i^N (p_i^+ + k_i^+) \Big)+ \cO(\lambda^4)
\,,\end{align}
where $\hat s = p_I^2$. We see that since $p_I$ is timelike, $1-z$ depends on the timelike component of the soft momentum outside of the jets and on the null component of the momentum within the jets.

So far, we have discussed the kinematics in the  hadronic endpoint defined as $\tau \to 1$, which is the main focus of this paper. However, $z$ can be forced close to 1 not only in this hadronic endpoint, but also in the limit of steeply falling parton luminosities. In this case, final states with small values of $\hat s$ are preferred, giving again $z \to 1$. Our analysis is independent of the precise mechanism which guarantees that $1-z$ can be regarded as a small quantity, and can therefore be used away from the true hadronic endpoint.

We are now in a position to discuss how our parametrization of $1-z$ reduces to the standard variable in the case of Drell-Yan and cases when there is one inclusive jet, such as $B \to X_s \gamma$, DIS, and direct gauge boson production. As we will see in \sec{factorization}, \eq{1-z} implies the soft function in general depends on the timelike component of $k_\out$ and on the null components (with respect to the corresponding jet directions) of the soft momenta in each of the jets $k_i$. In Drell-Yan, there are no jets in the final state and so the entire soft momentum $k_s$ is just $k_\out$. This is why the soft function in Drell-Yan depends only on the timelike component of the total soft momentum. For a single inclusive jet (\ie, defined with a jet algorithm that includes all of the hadronic momentum), all the soft momentum is included in the jet, such that $k_s = k_1$. This explains why the soft function in this latter type of process only depends on the null component of the total soft momentum. In Ref.~\cite{Kidonakis:1998bk}, on the other hand, threshold resummation for dijet production was considered and it was found that the soft function only depended on the timelike component of momentum outside of the jets. This apparent discrepancy is due to the fact that the limit of small jet size $R \to 0$ was taken and the contribution of in-jet soft particles vanishes in this limit.\footnote{Note that double counting is avoided in \cite{Kidonakis:1998bk} by removing collinear modes from the soft function (``eikonal subtractions''), which, for $R \sim \lambda$, removes any dependence on the soft momentum inside of jets, at leading order in the power counting.} From the discussion above, the soft function will have dependence on the null component of in-jet momentum for jets of finite size.

\section{Non-global Logarithms and the definition of threshold}
\label{sec:nonglobal}

In Ref.~\cite{Kidonakis:1998bk}, two definitions of $z$ were defined for the case of two final state jets, which we denote here as $z_a$ and $z_b$,
\begin{align}
&z_a \equiv \frac{(p_1+p_2)^2}{\hat s} \nn\\
&z_b \equiv \frac{2p_1\cdot p_2}{\hat s}
\,,\end{align}
where $p_{1,2}$ are the {\it total} 4-momenta of the jets . Note that, unlike $p_J$ (\cf~\eq{masslessPJ}), $p_{1,2}$ can not be defined by the jet direction and energy alone, so both definitions of $z$ are indirectly sensitive to the jet mass. To lowest order, $1-z_{a,b}$ can be written as
\begin{align}
1-z_a &= \frac{2 k_\out^0}{M_{JJ}} + \cO(\lambda^4) \nn\\
1-z_b &=\frac{2 k_\out^0}{M_{JJ}}  + \frac{p_1^2 + p_2^2}{M_{JJ}^2}  + \cO(\lambda^4)
\,,\end{align}
where on the right-hand side, $M_{JJ}^2$ can be set to $(p_1+p_2)^2$ for both $z_{a,b}$ to order $\lambda^4$.

It is well-known that jet observables which have an energy scale of radiation inside of a jet that is widely disparate from the scale outside the jet give rise to non-global logarithms \cite{Dasgupta:2001sh}. This is due to the fact that in this case, real emission corrections to radiation inside the jet can be vetoed when one of the daughter particles escapes the jet, and this leads to an incomplete cancellation of real and virtual corrections, an effect which is stronger for radiation near the jet boundary (the so-called ``buffer region"). As has been pointed out in the literature (see, \eg, Ref.~\cite{Banfi:2008qs}), $z_a$ introduces non-global logarithms, whereas $z_b$ does not. This is clear because the limit $1-z_a \ll 1$ only restricts radiation outside of the jet and, for a jet size $R \sim 1$, the radiation within the jet is not restricted in the hadronic endpoint $ \tau \to 1$. On the other hand, the scaling of in-jet and out-of-jet radiation is correlated with $z_b$ (in particular, in both cases soft radiation has momentum components that scale as $M_{JJ} (1-z_b)$) such that no non-global logarithms should arise.

Notice that for a two-jet final state in the hadronic endpoint ($\tau \to 1$), the definition of $z$ given in \eq{zdefn} reduces to 
\begin{align}
1-z \xrightarrow{\tau \to 1} \frac{2 k^0_{\out}}{M_{JJ}} +  2 \frac{p_1^2 + p_2^2}{M_{JJ}^2}   + \cO(\lambda^4)
\,,\end{align}
where we used \eq{1-z} and that, to $\cO(\lambda^2)$, $p_I = (\sqrt{\hat s}, \vec{0}) = (M_{JJ}, \vec{0})$ in the $\tau \to 1$ limit. Thus, our definition of $z$ restricts radiation both inside and outside of the jet, similarly to $z_b$ in this limit and should correspondingly not introduce non-global logarithms for $R \sim 1$. (This is also clear directly from \eq{1-z} which is valid away from the hadronic endpoint, provided another mechanism enforces $1-z \ll 1$  such as the steepness of PDFs.)

\section{Jet Algorithms at Hadron Colliders}
\label{sec:algorithm}

Perturbative calculations require a precise definition of the phase space boundaries imposed by the jet algorithms. Any jet algorithms needs to be infrared safe; otherwise, it leads to infrared divergent results when calculated in perturbation theory. There are two general types of jet algorithms: cone algorithms and cluster algorithms. Cone algorithms decide on which particles belong to a given jet based on cones of fixed size $R$, while cluster algorithms group particles together into jets based on a relative measure of their distance. These jet algorithms act on the entire set of particles in the final state to decide how many jets are contained in a given event and which particles belong to which jet. Almost all jet algorithms depend on a jet size $R$, and a distance $\jetR_{ij}$ that measures the distance between two particles in $\eta-\phi$ space
\begin{equation}
\label{eq:pp-measure}
\jetR_{ij} = \sqrt{(\Delta \eta_{ij})^2 + (\Delta \phi_{ij})^2}
\,.\end{equation}

As already discussed in the previous section, the relevant degrees of freedom in jet production close to $z=1$ are collinear and soft particles. To perform perturbative calculations in this region we therefore need a restriction on these degrees of freedom to decide whether they belong to a given jet or not. Collinear particles in a given direction all belong to the same jet. This is in contrast with soft particles, which can either belong to a jet or not. Note that the treatment of jet algorithms in SCET is only correct to leading order in the power counting parameter $\lambda$. Therefore, we assume that all jets have energy much in excess of their mass, and that all jets are widely separated. 

Standard cone algorithms, such as SISCone, are quite simple. The restrictions they impose on each particle to belong to a given jet are independent of other particles in the event, and only depend on the angular distance from the jet direction. The restriction for both soft and collinear particles to be in a jet $j$ with direction $n$ is therefore
\begin{equation}
\label{eq:theta_general_cone}
\hat \Theta^{R}_{{\rm soft},{j}}  = \hat \Theta^{R}_{{\rm coll},{j}}  = \prod_i \Theta(\jetR_{i,{n}} < R)\,.
\end{equation}
For the purposes of this paper, we only need results at relative order $\alpha_s$, and therefore only have to consider one extra particle in the final state. The restrictions therefore simplify, and for the extra particle $i$ we can write
\begin{equation}
\hat \Theta^{R}_{{\rm soft},{j}}  = \hat \Theta^{R}_{{\rm coll},{j}}  = \Theta(\jetR_{i,{n}} < R)\,.
\end{equation}

Cluster algorithms iterate a process of calculating a distance measure $d_{ij}$ for all pairs of particles and a distance $d_i$ for each jet and, if a $d_i$ is smallest, removing particle $i$, or, if a $d_{ij}$ is smallest, merging jets $i$ and $j$. In general , $d_{ij}$ is defined as $d_{ij} = {\rm min}\{d_i, d_j\} \jetR_{ij}/D$, where $D$ is a fixed parameter that characterizes the size of a jet and the precise definition of $d_{i}$ depends on the choice of algorithm.\footnote{An example for such a distance measure is $d_{i} = p_{\T}^i$ for the $k_\T$ algorithm.} This makes the action of the jet algorithm considerably more complicated. As explained above, all collinear particles in a given direction have to end up in the same jet, which allows us to write a generic restriction for the action of a cluster algorithm on a set of collinear particles in a given direction as
\begin{equation}
\hat \Theta^{R}_{{\rm coll},{j}}   = \prod_{k=0}^{N-1}\Theta(\jetR^{N-k}_{{\rm min}} <R)
\,.\end{equation}
Here, $N$ denotes the total number of collinear particles in the direction of the jet $n_j$. $\jetR^{N-k}_{{\rm min}}$ denotes the $\jetR_{ij}$ between the pair of collinear particles in the set of $N-k$ remaining particles with the smallest $d_{ij}$. For soft particles, such a generic formula is not possible (at least analytically), since different soft particles can end up in different jets, and the restriction on a given particle depends on all other soft particles in the event. At relative order $\alpha_s$, however, the restrictions $\Theta^R_{i}$ for cluster algorithms simplify and are given by
\begin{align}
\hat{\Theta}^R_{{\rm soft},{j}}=\Theta(\jetR_{i,{n}} < R)
\end{align}
for soft particles and by
\begin{align}
\hat{\Theta}^R_{{\rm coll},{j}}=\Theta(\jetR_{k,{l}} < R)
\end{align}
for collinear particles, where $k$ and $l$ label the new particles after the collinear splitting. For more information about jet algorithms in SCET, see \cite{Ellis:2010rw}.

\section{$N$-Jet Factorization Theorem}
\label{sec:factorization}

In this section, we present the factorization theorem for the cross section to produce $N$ jets, defined with respect to a jet algorithm, differential in the 3-momentum ($p_\T$ and pseudorapidity $\eta$) of each jet and of the non-strongly interacting particles. To keep the notation simple, we begin in \subsec{1jet} by discussing the case of a single jet produced via the channel $q\bar{q} \to g$. We then discuss the differences between this derivation and the one needed for the channel $qg \to q$. It will be clear from these derivations that, aside from the promotion of the hard and soft functions to matrices which arise from mixing of operators in color space, there is nothing conceptually or technically new for arbitrary $N$-jet production in our approach. This allows us to generalize our results to the $N$-jet factorization formula in \subsec{Njet}.

In writing down a factorization theorem, we first assume that we can match QCD onto operators in SCET containing $N+2$ distinct collinear fields. This is valid when a (direct or indirect) measurement constrains the final state to be $N$-jet like. In our case, this is ensured by the fact that we take the variable $1-z$ to be small, together with the assumption that the jets are well separated from each other and from the beams (with the latter requirement ensuring that the probability of initial-state collinear radiation to produce a jet is power suppressed relative to the probability of the jet arising from the hard interaction). 
Our derivation is agnostic as to the cause of $1 - z  \ll 1$, and in \cite{threshold2} we explore in greater detail in what regimes the steepness of parton luminosities allow the factorization theorem we derive here to be applied away from the hadronic endpoint.
We will assume in this section that the reader has some familiarity with SCET. For details, we refer the reader to the original SCET literature~\cite{Bauer:2000ew, Bauer:2000yr, Bauer:2001ct, Bauer:2001yt}.
\subsection{Case of a Single Jet}
\label{subsec:1jet}

\subsubsection{$q \bar q \to g$} 

Working to leading order in the electroweak coupling constant, we first write the full theory matrix element mediating the partonic interaction as
\begin{equation}
\label{eq:MfullQCD}
\mae{qX}{O}{P_1 P_2} = \sum_i M_i^{\alpha \beta \mu} T^A_{ab} \, \mae{X}{\bar{\psi}^\alpha_a \psi^\beta_b A_\mu^A}{P_1 P_2}
\,.\end{equation} 
Here, $\ket{q}$ represents the non-strongly interacting final state of total momentum $q$, $\ket{P_1}$ and $\ket{P_2}$ are the incoming hadrons with the corresponding momentum, and $\ket{X}$ represents the hadronic final state. The index $i$ labels Dirac structures. This equation defines the $M_i^{\alpha \beta \mu}$. Note that we have used the fact that there is only one color singlet in the decomposition of $3 \otimes \bar{3} \otimes 8$.

In terms of $M_i$, the matching of QCD onto the fields of SCET takes the form
\begin{align}
\label{eq:3jetmatch}
 M_i^{\alpha \beta \mu} \cQ^{\alpha \beta \mu}(x) &\equiv M_i^{\alpha \beta \mu} \big[ \bar{\psi}^\alpha_a \psi^\beta_b A_\mu^A \big](x) \nn\\
 &= \sum_j M_j^{\alpha \beta \mu} \sum_{\{\ti{p}\}} C_{ij}(\{\ti{p}\}) e^{i (\ti{p}_1+\ti{p}_2 - \ti{p}_3) \cdot x} \big[ (\bar{\chi}_{- \ti{p}_1})^\alpha_a (\chi_{\ti{p}_2})^\beta_b (B_{-\ti{p}_3})_\mu^A \big](x)\,,
\end{align}
where $\ti{p}_i$ is the label momentum carried by the field. At tree level, we have
\begin{equation}
C_{ij}(\{\ti{p}\}) =  \delta_{ij}
 \,.\end{equation}
The matching condition in momentum space takes the form
 \begin{align}
 \label{eq:match}
 M_i^{\alpha \beta \mu} \cQ^{\alpha \beta \mu}(k) &\equiv M_i^{\alpha \beta \mu} \int\!\df^4x \,e^{-ik\cdot x} \cQ^{\alpha \beta \mu}(x) \nn\\
&=  \sum_jM_j^{\alpha \beta \mu} \bigg(\prod_{i=1}^3 \int\! \dfslash ^4 p_i \bigg)  C_{ij}(\{\ti{p}\}) \big[\bar{\chi}^\beta_b(- p_1) \, \chi^\alpha_a(p_2)\, B_\mu^A(-p_3)\big]  \nn
\\ & \qquad \times \deltaslash^4(p_1+p_2-p_3 - k)\,,
\end{align}
where we turned the sums over labels and integrals over residual momenta into integrals over the full $\df^4 p_i$ and used the shorthand notation
\begin{equation}
\int\!\dfslash ^4 p \equiv \int\!\frac{\df^4 p}{(2 \pi)^4}
\,.\end{equation}

Using this matching condition, we can write the cross section differential in the jet pseudorapidity $\eta_J$ and transverse momentum $p_J^\T$ as 
\begin{align}
\label{eq:fact1}
\frac{\df \sigma}{\df^2 p^\T_J \df \text{tanh}\thinspace\eta_J \df \Phi_q} &= \frac{1}{2 \Ecm^2} \sum_X^{\rm rest.} \vert \mae{qX}{O}{P_1 P_2} \vert^2_{\text {spin avg.}} \delta(\eta_J - \eta(X)) \delta^2(\vec{p}_J^\T - \vec{p}^\T(X)) \\
&\quad \times (2 \pi)^4 \delta^4(P_1 + P_2 - q - p_X) \nn\\
&= \frac{1}{2 \Ecm^2} \sum_{X}^{\rm rest.}\sum_{\substack{i,j ,i',j',\\\rm spin}} M_j^{\alpha \beta \mu} \overline{M}_{j'}^{\bar \beta \bar \alpha \bar \mu} \frac{T^A_{ab}T^{\bar A}_{\bar b \bar a}}{4 \CA^2} \Big( \prod_{k=1}^3 \int\, \dfslash^4 p_k  \dfslash^4 p_k' \Big) C_{ij}(\{\ti{p_k}\}) C_{i'j'}^*(\{\ti{p}_k'\}) \nn\\
& \quad \times  \mae{P_1 P_2}{\bar{\chi}^{\bar \beta}_{\bar b}(p_2') \, \chi^{\bar \alpha}_{\bar a}(-p_1') B_{\bar \mu}^{\dagger\bar A}(-p_3')}{X}\mae{X}{\bar{\chi}^\alpha_a(- p_1) \, \chi^\beta_b(p_2) B_\mu^A(-p_3)}{P_1 P_2} \nn\\
& \quad \times  \delta(\eta_J - \eta(X)) \delta^2(\vec{p}_J^\perp - \vec{p}^\perp(X)) \, \deltaslash^4(p_1+p_2 -p_3-q) \nn\,.\end{align}
Here, we defined $\df \Phi_q$ as the phase space measure of the $m$ non-strongly interacting final-state particles,
\begin{equation}
\df \Phi_q \equiv \prod_k^m \frac{\dfslash^3 q_k}{2 E_k}\,,
\end{equation}
where we have omitted symmetry factors coming from identical particles in $\Phi_q$.
The restriction on the sum over final states $X$ (``rest.'') is that they include exactly one jet as defined by the jet algorithm and the delta functions that fix $\eta(X)$ and $\vec{p}^\T(X)$ act on the part of $X$ identified to be the jet, which we assume to be sufficiently separated from the beams such that contributions from collinear initial-state radiation are power suppressed. We also define $\overline M \equiv \gamma^0 M^\dagger \gamma^0$. To arrive at this equation, we used the hadronic momentum-conserving delta function in the first line to shift the operator $O^\dagger$ to the point $x$, applied the matching condition \eq{match}, and then integrated over $x$ which resulted in the partonic momentum-conserving delta function on the last line.

We can simplify \eq{fact1} using the following observations. First, we can use the BPS field redefinition \cite{Bauer:2001yt} to decouple soft and collinear modes to $\cO(\lambda^2)$,
\begin{subequations}
\label{eq:BPS}
\begin{align}
\chi_{n}(x) & \to Y_{n}(x)\chi_{n}(x) \\
\bar\chi_{n}(x) &\to \bar\chi_{n}(x){Y}^\dag_{n}(x) \\
B_{n}(x) &\to \mathcal{Y}_n(x) B_{n}(x) =Y^\dagger_n(x) B_{n}(x) Y_n(x)
\,,\end{align}
\end{subequations}
where $Y_n$ is a soft Wilson line, for which we adopt the conventions\footnote{For a discussion of the various conventions for incoming and outgoing Wilson lines and how they are related see, for example, Ref.~\cite{Arnesen:2005nk}. There is also a nice discussion of how soft Wilson lines arise in the path integral formulation of SCET in Appendix C of Ref.~\cite{Becher:2007ty}.}
\begin{subequations}
\label{eq:Yout-in}
\begin{align}
\label{eq:Yin}
Y^{\ins}_{n}(x) = P\, \exp\biggl[ \img g_s \int_{-\infty}^0\!\df s \, n \cdot A_s(x+sn)\biggr]
\,,\end{align}
for an incoming particle (where $P$ denotes path ordering) and
\begin{align}
\label{eq:Yout}
Y_{n}^{\out \, \dagger}(x) = P\exp\biggl[\img g_s \int^{\infty}_0\!\df s \, n \cdot A_s(x+s\,n)\biggr]
\,,\end{align}
\end{subequations}
for an outgoing particle. In momentum space, the field redefinition in \eq{BPS} induces the shift $p_c \to p_c + k_s$, where $k_s$ is the total soft momentum carried by the Wilson lines. 

Second, we can implement the restriction on the final state $X$ that there is one jet at the operator level to all orders by using the jet algorithm operator $\hat \Theta^R_i$, defined in \sec{algorithm}. This allows us to complete the sum over $X$ and factorize the collinear and soft matrix elements from one another.

The final state collinear matrix element can be written as
\begin{align} 
\label{eq:Jg_generic}
& \Mae{0}{B^{\dagger A}_{\mu}(-p_3')\, \hat \Theta_{{\rm coll},3}^R \delta(\eta_J - \hat \eta) \delta^2(\vec{p}_J^\T - \hat{\vec{p}}^\T) \, B^{B}_\nu(-p_3)}{0}  =  - E_J \deltaslash^4(p_3' - p_3)\delta^{A B} g^{\perp_3}_{\mu \nu} J^g_\omega(p_3)
\,,\end{align}
which defines the gluon jet function $J_w^g(p)$. Integrating \eq{Jg_generic} over the full $p_3$ and $p_3'$, which contain integrals over residual momentum and sums over the labels $\omega$ and $n_3$ of the $B_\perp$ field, fixes the labels to be $n_3 = (1, \vec{p}_J^\T/\abs{\vec{p}_J^\T} \cosh \eta_J, \tanh \eta_J)$ and $\omega = 2 p_J^\T \cosh \eta_J$. The label ``3'' on $\hat \Theta_R^3$ and on $g^{\perp_3}_{\mu \nu}$ indicates these are defined with respect to the direction $n_3$ of the jet. Note that this jet function depends on the choice of the jet algorithm. 

For an algorithm that is inclusive over collinear initial-state radiation, the initial-state collinear matrix elements give rise to PDFs from the relations  \cite{Bauer:2002nz,Bauer:2008jx}
\begin{align}
\label{eq:PDF}
\int\!\dfslash^4 p \,\dfslash^4 p' \mae{P}{\bar{\chi}^{\alpha'}_{a'}(p') \chi^\alpha_a(p)}{P}_{\text{spin avg.}} &= \Ecm \delta_{a a'} \Big(\frac{\nslash}{2} \Big )^{\alpha \alpha'} \int_0^1 \df x f_q(x) \nn\\
\int\!\dfslash^4 p \,\dfslash^4 p' \mae{P}{\chi^{\alpha'}_{a'}(-p') \bar{\chi}^{\alpha}_{a}(-p)}{P}_{\text{spin avg.}} &=  \Ecm \delta_{a a'} \Big(\frac{\bnslash}{2} \Big )^{\alpha' \alpha} \int_0^1 \df x f_{\bar q}(x)
\,,\end{align}
with $p_{1,2}$ set to $x_{1,2}\Ecm \frac{n_{1,2}}{2}=\frac{1}{2}\omega_{1,2}\ts n_{1,2}$ (where $n_1 \equiv n$ and $n_2 \equiv \bar n$) and $p_{1,2}' = p_{1,2}$.

The color structure in the matrix elements \eqs{Jg_generic}{PDF} leads to a trace over the color structure of the Wilson lines in the soft function.
We can write the soft function in terms of the variables of interest $k_\out$ and $k_3$ (the out-of-jet and in-jet momenta) as
\begin{align}
\label{eq:3jetsoft}
\int\!\dfslash^4 k_s S(k_s, \{ n_i\})
&=  \frac{1}{\CA \CF}  \!\int\!\!\df^4 k_s\Mae{0}{ \overline{\T} \Big[  Y_{n_2}^{\dagger } Y_{n_3}^\dagger T^A Y_{n_3} Y_{n_1}\Big]  \delta^4(k^\mu_s - i\partial^\mu)   \T \Big[  Y_{n_1}^\dagger Y^\dagger_{n_3} T^A Y_{n_3} Y_{n_2}\Big]}{ 0}\nn\\
& =   \int\! \df^4 k_\out \, \df^4 k_3 \, S(k_\out, k_3, \{ n_i\})
\,,\end{align}
where $\{ n_i\} = \{n_q, n_{\bar q}, n_g \}$ and $\T$ ($\overline{\T}$) denotes (anti-) time ordering. $S(k_\out, k_3, \{ n_i\})$ is then defined as
\begin{align}
S(k_\out, k_3, \{ n_i \}) &\equiv  \frac{1}{\CA \CF} \Mae{0}{ \overline{\T}\!\Big[  Y_{n_2}^{\dagger } Y_{n_3}^\dagger T^A Y_{n_3} Y_{n_1}\Big]  \delta^4(k_3 -\hat k_3) \nn\\
& \qquad \qquad \times \delta^4(k_\out -\hat k_\out)  \T\!\Big[  Y_{n_1}^\dagger Y^\dagger_{n_3} T^A Y_{n_3} Y_{n_2}\Big] }{0}\,,
\end{align}
and it is understood that the replacement $k_s \to k_\out + k_3$ should be made wherever $k_s$ appears. The operators $\hat k_\out$ and $\hat k_3$ are defined as
\begin{align}
\label{eq:k_inout}
\hat k^\mu_3 &\equiv \hat \Theta_{{\rm soft},3}^R  i \partial^\mu \nn\\
\hat k^\mu_\out &\equiv \Big( 1 - \hat \Theta_{{\rm soft},3}^R \Big) i \partial^\mu
\,.\end{align}

Finally, using that the spin- and color-averaged square of the Born matrix element $M_B$, $\overline{\abs{M_B}^2}$, can be written in terms of $M_i$ as
\begin{align}
\label{eq:born-ME}
\overline{\abs{M_B}^2} \equiv \frac{ 1}{4 \CA^2}\sum_{\rm spin, color}  \vert M_B \vert ^2 &= - \frac{\Tr[T^A T^A]}{4 \CA^2} \sum_{i,i'} M_i^{\alpha \beta \mu} \overline{M}_{i'}^{\bar \alpha \bar \beta \bar \mu} (\dslash{p_1})^{\alpha \bar \alpha} (\dslash{p_2})^{\bar \beta \beta} g_{\mu \bar \mu}^{\perp_3} \nn\\
&= - \frac{\CF}{4 \CA} \hat s \sum_{i,i'} M_i^{\alpha \beta \mu} \overline{M}_{i'}^{\bar \alpha \bar \beta \bar \mu} \Big(\frac{\nslash_1}{2} \Big)^{\alpha \bar \alpha} \Big(\frac{\nslash_2}{2} \Big)^{\bar \beta \beta} g_{\mu \bar \mu}^{\perp_3}
\,,
\end{align}
we see that the Dirac structure of the matrix elements in \eqs{Jg_generic}{PDF} naturally gives rise to the Born cross section. 

To simplify the notation, we define a hard function $H$ which includes all perturbative corrections contained in the matching coefficients $C_{ij}$ as
\begin{equation}
H( \{ n_i , \omega_i\}) \equiv -\frac{\frac{\CF}{4 \CA} \hat s \sum_{i,i',j,j'} C_{ij}(\{\ti{p}_k\}) C_{i'j'}^*(\{\ti{p}_k\}) M_j^{\alpha \beta \mu} \overline{M}_{j'}^{\bar \beta \bar \alpha \bar \mu} \Big(\frac{\nslash_1}{2} \Big)^{\alpha \bar \alpha} \Big(\frac{\nslash_2}{2} \Big)^{\bar \beta \beta} g_{\mu \bar \mu}^{\perp_3}}{\overline{\vert M_B\vert ^2} }
\,,\end{equation}
where $\omega_i$ are the labels on the three collinear fields. 
$H$ by definition is $1$ to leading order in $\alpha_s$. Putting this together, we arrive at the expression
\begin{align}
\label{eq:fact2}
\frac{\df \sigma}{\df^2 p^\T_J \df \text{tanh}\thinspace\eta_J \df \Phi_q} &=\frac{E_J}{2\Ecm^2} \int\!\frac{\df x_1}{x_1}\frac{\df x_2}{x_2} \overline{\vert M_B\vert^2} \int\!\dfslash^4 p_3 \int\!\df^4 k_\out  \int\!\df^4 k_3   \\
& \,\,\, \times H( \{ n_i , \omega_i\}) f_q(x_1) f_{\bar q}(x_2)  J_\omega^g(p_3) S_{q \bar q \to g}(k_\out, k_3, \{ n_i \}) \, \nn\\
& \,\,\, \times \deltaslash^4(x_1 \Ecm \frac{n}{2} + x_2 \Ecm \frac{\bar n}{2} -q -k_\out - p_3 - k_3) \nn
\,.\end{align}

The final step is to simplify the momentum-conserving delta function. We use \eqs{p-decomp}{p-cons2} to write it, up to power corrections in $\lambda$, as
\begin{align}
\label{eq:p-cons-simp}
& \delta^4(x_1 \Ecm \frac{n}{2} + x_2 \Ecm \frac{\bar n}{2}-q -k_\out - p_3 - k_3)   \nn\\
&\qquad \qquad =   \frac{2}{\Ecm^2 \tau} \delta \Big(1-z -   \frac{1}{\hat{s}} \Big[ 2 p_I \cdot k_\out + 2 p_I^0( p_c^+ + k_3^+)\Big]  \Big) \nn\\
&  \qquad \qquad \qquad \times \delta^2(\vec{p}_J^\T + \vec{q}_J^\T) \, \delta\!\left( Y - \arctanh \left( \frac{p_J^\T \sinh \eta_J + q_z}{p_J^\T \cosh \eta_J + q_0}  \right) \right) 
\,,\end{align}
where $p_I$ is the total (partonic) initial-state momentum, $p_I^ 0 = (x_1 + x_2) \Ecm/2$ and $p_I^z = (x_1 - x_2) \Ecm/2$. Here, we used that
\begin{equation}
\delta\bigl[f(x,y)\bigr]\delta\bigl[g(x,y)\bigr]=\frac{\delta(x-x_0)\delta(y-y_0)}{\left\vert\frac{\partial (f, g) }{\partial (x, y) }\right\vert}\,,
\end{equation}
where $f(x_0,y_0)=g(x_0,y_0)=0$ and made the change of variables
\begin{align}
\label{eq:x1x2-to-zY}
x_1 & = \sqrt{\frac{\tau}{z}} e^Y \nn\\
x_2 & = \sqrt{\frac{\tau}{z}} e^{-Y}
\,.\end{align}
In switching from $x_{1,2}$ to $z$ and the total rapidity $Y$, we will also need that 
\begin{align}
\int\!\frac{\df x_1}{x_1} \frac{\df x_2}{x_2} = \int_\tau^1\!\frac{\df z}{z} \int_{-\frac{1}{2}\ln \frac{z}{\tau}}^{\frac{1}{2}\ln\frac{z}{\tau}}\df Y\,.
\end{align}
Now, since there is no dependence on the soft momenta other than on the components $p_I \cdot k_\out$ and $k_3^+$ and the only unconstrained component of $p_3$ is the plus-component $p_3^+$, we can integrate over the other variables to obtain our final expression
\begin{align}
\label{eq:qqg-fact-result}
\frac{\df \sigma}{\df^2 p^\T_J \df \text{tanh}\thinspace\eta_J \df \Phi_q} &=   \frac{\pi}{\Ecm^4 \tau} \int_\tau^1\!\df z \int_{-\frac{1}{2}\ln \frac{z}{\tau}}^{\frac{1}{2}\ln\frac{z}{\tau}}\df Y\,\overline{\vert M_B\vert^2}  H_{q \bar q \to gX}( \{ n_i , \omega_i\}) f_q(x_1) f_{\bar q}(x_2)   \nn\\
& \quad  \times \int\!\df p_{3}^+  J_\omega^g(p_3^+)  \int\!\df  k_{\out}^0 \int\! \df k_{3}^+S_{q \bar q \to g}( k_{\out}^0, k_{3}^+)  \delta^2(\vec p_J^\T +\vec q^\T) \nn\\
& \quad \times \delta\!\left( Y - \arctanh \left( \frac{p_J^\T \sinh \eta_J + q_z}{p_J^\T \cosh \eta_J + q_0}  \right) \right) 
 \nn\\
& \quad \times  \delta \left[1-z - \frac{2}{\sqrt{\hat s}} \Bigl(\cosh Y   \big(p_{c }^+ + k_{\ins }^+ \big)  +   k^0_{\out} \Bigr)\right]
\,.\end{align} 
The soft function in \eq{qqg-fact-result} is defined as
\begin{align}
S_{q \bar q \to g}(k_\out^0, k_3^+,  \{ n_i\}) &\equiv  \frac{1}{\CA \CF} \Mae{0}{\overline{\T}\!\Big[  Y_{n_2}^{\dagger } Y_{n_3}^\dagger T^A Y_{n_3} Y_{n_1}\Big] \delta(k_3^+ -\hat k^+_3) \delta\Big(k_\out^0 -\frac{ p_I \cdot \hat k_\out}{\vert p_I \vert} \Big)  \nn\\
& \qquad \qquad  \qquad \times \T\!\Big[  Y_{n_1}^\dagger Y^\dagger_{n_3} T^A Y_{n_3} Y_{n_2}\Big]  }{0} \nn\\
&= \delta(k_\out^0) \delta(k_3^+) + \cO(\as)
\,.\end{align}
Note that, despite our notation, $k_\out^0 \equiv p_I \cdot k_\out/ \vert p_I\vert $ is simply the projection of the soft momentum outside of all jets onto a timelike vector. This function is most easily computed in the partonic center-of-mass frame where $p_I$ is in fact purely timelike. However, in general $p_I$ will have nonzero spatial components, in which case $k_{\out}^0$ is not simply the energy component of the $k_{\out}$ four-vector.
The jet function in \eq{qqg-fact-result} is defined as
\begin{align}
\label{eq:Jg}
 J^g_\omega(p^+) g_\perp^{\mu \nu}\delta^{AB}\equiv - \frac{\omega}{2\pi} \int\!\df^4 x \, e^{i p\cdot x} \bra{0} B_{\perp, \omega}^{\mu,A} (x) \hat \Theta_{{\rm coll},3}^R B_{\perp, \omega}^{\nu,B}(0)\ket{0}  = \delta(p^+) g_\perp^{\mu \nu}\delta^{AB}+ \cO(\as)
\,,\end{align}
where, again, the label $\omega$ is set to $\omega = 2 E_J  = 2 p^J_\T \cosh \eta_J$.

\subsubsection{$q g \to q$} 

The majority of the above discussion goes through in much the same way for the channel $qg \to q$. The main differences are that the matrix element of final state fields gives rise to a quark jet function, defined by
\begin{align}
\label{eq:Jq_generic}
& \int\!\dfslash^4 p_3'\,\Mae{0}{\chi^{c'}_{\alpha'}(p_3')\, \hat \Theta_{{\rm coll},3}^R \delta(\eta_J - \hat \eta) \delta^2(\vec{p}_J^\T - \hat{\vec{p}}^\T)
   \bar\chi^c_\alpha(p_3)}{0}
= \delta^{c'c} \Bigl(\frac{\dslash{n}_3}{2}\Bigr)_{\alpha'\alpha} J_\omega^q(p_3)
\,,\end{align}
which, after integrating over all but the $p^+$ component, becomes
\begin{align}
\label{eq:Jq}
 J^q_\omega(p^+)\delta^{ab} \equiv \frac{1}{2\pi} \int\!\df^4 x \, e^{i p\cdot x} \bra{0} \frac{\bnslash}{2} \chi_{n, \omega}^a(x)   \hat \Theta_{{\rm coll},3}^R  \bar\chi_{n, \omega}^b (0)\ket{0}  = \delta(p^+)\delta^{ab} + \cO(\as)
\,.\end{align}
For the initial-state gluon, we obtain the gluon PDF via the relation
\begin{align}
\int\!\dfslash^4 p \ts\dfslash^4 p' \mae{P}{ B^{\dagger A}_\nu(p') B^B_\mu(p)}{P} &= \frac{g_{\mu\nu}^\perp}{D-2} \delta_{AB} \int_0^1  \frac{\df x}{x} f_g(x) 
\,.\end{align}
Finally, the soft function is defined as in \eq{3jetsoft} but with the appropriate modification in the definition of the Wilson lines for incoming and outgoing fields given in \eq{Yout-in}.
The final result after these differences are taken into account is of the form \eq{qqg-fact-result} but with the substitutions $f_{\bar q} \to f_g$, $J^g \to J^q$, and $S_{q \bar q \to g} \to S_{qg \to q}$.

\subsection{Extension to $N$ Jets}
\label{subsec:Njet}

The above results clearly generalize. The only nontrivial complication is due to mixing in color space. To avoid cumbersome notation, we will explain what generalizes and state the result rather than write down the $N$-jet derivation.  Explicitly, for every quark, antiquark and gluon in the final (initial) state, the all-orders jet function (PDF) has precisely the Dirac and color structure to contract with the non-QCD matrix element $ M_i^{\alpha \beta \cdots \mu \cdots}$ to give the Born matrix element at tree level. The PDFs and jet functions have the same definitions as in the single jet case. However, due to the fact that operators with different color structures in general mix, the hard and soft functions in these formulas should be interpreted as matrices in color space.

Since the main difference in the generalization to $N$ jets is in the soft function, we will discuss it in more detail. 
It takes the general form
\begin{align}
\label{eq:soft-generic}
S(k_\out^0, \{n_i ,k_i^+\}) \equiv \frac{1}{\mathcal{N}} \bra{0} O_S^\dagger \, \delta\Big(k_\out^0 -\frac{ p_I \cdot \hat k_\out}{\vert p_I \vert} \Big)  \prod_i^N \delta (k_i^+ - \hat k_i^+) \, O_S \ket{0}
\,,\end{align}
where $\mathcal{N}$ is a normalization factor such that the soft function is unity (times delta functions in its arguments) at tree level and the operator $O_S$ is the product of Wilson lines that arise from the BPS field redefinitions \eq{BPS}, appropriately traced. As for the one jet case, the arguments $k_i$ in the soft function run over all final state jets and the $n_i$ include all directions, both initial and final. The operators $\hat k_\out$ and $ \hat k_i$ that appear in \eq{soft-generic} are defined as
\begin{align}
\label{eq:k_inout-generic}
\hat k^\mu_i &\equiv \hat \Theta_{{\rm soft},i}^R  \, i \partial^\mu \nn\\
\hat k^\mu_\out &\equiv \Big( 1 - \sum_i^N \hat \Theta_{{\rm soft},i}^R \Big) i \partial^\mu
\,.\end{align}
The result of going through the same steps as for the single jet case leads to the $N$-jet factorization formula,
\begin{align}
\label{eq:Njet-fact-result}
\frac{\df \sigma}{\prod_i^N \df^2 p^\T_{J_i} \df \text{tanh}\thinspace\eta_{J_i} \df \Phi_q} &=  \frac{1}{\Ecm^4 \tau} \int_\tau^1\!\df z \int_{-\frac{1}{2}\ln \frac{z}{\tau}}^{\frac{1}{2}\ln\frac{z}{\tau}}\df Y\,  \overline{\vert M_B\vert^2}  H( \{ n_i, \omega_i\}) f_1(x_1) f_{ 2}(x_2)  \nn\\
& \quad  \times   \prod_i^N \int\!\frac{\df p_{i}^+}{2\!\cdot\!(2\pi)^3} J_i(p_i^+)  \prod_i^N  \int\!\df k_{i}^+ \int\!\df  k_{\out}^0  S( k_{\out}^0, \{  n_i ,k_{i}^+\}) \, 
 \nn\\
& \quad \times  (2\pi)^4\thinspace\delta \left[1-z - \frac{2}{\sqrt{\hat s}} \Bigl(\cosh Y  \sum_i^N \big(p_{i }^+ + k_{i }^+ \big)  +   k^0_{\out} \Bigr)\right] \nn\\
 & \quad \times \delta\!\left( Y - \arctanh \left( \frac{\sum_i^N p_{J_i}^\perp \sinh \eta_{J_i} + q_z}{\sum_i^N p_{J_i}^\perp \cosh \eta_{J_i} + q_0}  \right) \right) \delta^2(\sum_i^N \vec p_{J_i}^\T +  \vec q^\T) 
\,.\end{align}
In \eq{Njet-fact-result}, both $H$ and $S$ are matrices in color space, while $J$ and $f$ are proportional to the identity in color space and there is an implicit trace. Loop corrections will in general mix color structures, which means that beyond tree level $H$ and $S$ will contain off diagonal elements. The details of resummation in the presence of color mixing is discussed in \cite{Bonciani:2003nt}.
\section{Anomalous dimensions}
\label{sec:NLOresults}

Now that we have shown that a generic $N$-jet cross section factorizes in the limit of $1-z \ll 1$, we are left to calculate each ingredient of the factorization theorem. In this paper, we focus on the consistency of the factorization theorem to $\cO(\as)$, and we therefore need the one-loop anomalous dimensions for the hard, jet and soft functions. Note that the results presented here are enough to resum threshold logarithms at NLL\footnote{There are several different ways to define precisely what is meant by NLL. In this paper, we will use the convention of \cite{Becher:2007ty}.}.
\subsection{Hard, Jet, and Parton Distribution Functions}
Both the Born-level matrix element and the hard function, which can be found by calculating the virtual corrections to the Born-level matrix element in the $\overline{\rm MS}$ scheme, are process dependent, so they can not be calculated generically. However, the hard anomalous dimension, defined as
\begin{equation}
\frac{\df H(\{n_i,\omega_i\}; \mu)}{\df \, \ln \mu} \equiv \gamma_H\left(\{n_i,\omega_i\}; \mu\right) H\left(\{n_i,\omega_i\}; \mu\right)\,,
\end{equation}
only depends on the directions $n_i$, label momenta $\omega_i$, and color charges $\vec{T}_i$ of the collinear particles and is given by (\cite{Chiu:2009mg}, \cite{Aybat:2006mz}, \cite{Becher:2009cu}, \cite{Gardi:2009qi}, \cite{Becher:2009qa})
\begin{align}
\label{eq:gamma_H_res}
\gamma_H(\{n_i,\omega_i\}; \mu)=\sum_{i\in\partset} \left(\!-\Gamma_{\cusp}\ts \vec{T}_i^2\ts\ln\frac{\mu^2}{\omega_{i}^2}-\gamma_i \right)-2\ts\Gamma_{\cusp}\sum_{\pair{i}{j}}\vec{T}_i\mcdot \vec{T}_j\ts\ln\frac{n_i\mcdot n_j}{2}\,.
\end{align}
The cusp anomalous dimension at two loops is given by
\begin{equation}
\Gamma_{\rm cusp} = \frac{\alpha_s}{\pi}  + \frac{\alpha_s^2}{4 \pi^2} \left[ C_A \left( \frac{67}{9} - \frac{\pi^2}{3} \right) - \frac{20}{9} C_F T_F n_f \right]
\,.\end{equation}
The $\gamma_i$ to one-loop are given by
\begin{equation}
\gamma_q=\frac{3\as}{2\pi}\CF\qquad\text{and}\qquad\gamma_g=\frac{\as}{2\pi}\beta_0\,,
\end{equation}
for quarks and gluons, respectively, where $\beta_0$ is defined as
\begin{align}
\beta_0=\frac{11C_A}{3} - \frac{2\NF}{3}\,.
\end{align}
The sum on $\pair{i}{j}$ is a sum over all distinct pairs of partons $i$ and $j$ for $i\neq j$.

The quark and gluon jet functions have been calculated previously, \eg \cite{Bauer:2003pi, Bosch:2004th,Bauer:2001rh}, and were first calculated with a jet algorithm in \cite{Ellis:2010rw}. Their anomalous dimensions, defined by
\begin{equation}
\frac{\df J_i(p_i^+; \mu)}{\df \,\ln \mu} = \int_0^{p_i^+} \df {p'}_i^+\gamma_{J_i}({p}_i^+-{p'}_i^+; \mu) J_i({p'}_i^+; \mu)\,,
\end{equation}
are given by
\begin{align}
\label{eq:gamma_J_res}
\gamma_{J_i}(p_i^+; \mu)=\Big(2\ts\Gamma_{\cusp}\ts \vec{T}_i^2\ts\ln\frac{\mu}{\omega_i} + \gamma_i \Big)\ts\delta(p_i^+)-2\ts\Gamma_{\cusp}\ts \vec{T}_i^2\ts\frac{1}{\mu}\left(\frac{\mu}{p_i^+}\right)_{\plus}\,.
\end{align}
The expressions for $\Gamma_{\rm cusp}$ and $\gamma_i$ are the same as for the hard function. Note that the algorithm in \cite{Ellis:2010rw} used a polar angle for the measure and not \eq{pp-measure}. However, since the anomalous dimension in that case did not depend on the algorithm parameter $R$, the result must be independent of which measure is chosen since the precise definition of the jet boundaries is not associated with any singularities. It does however affect the finite parts of the jet function, which become important starting at NNLL accuracy.

It is well-known that the parton distribution functions are not perturbatively calculable; in practice, they can be expressed as universal matrix elements, which are then extracted from experiment. However, the evolution of the PDFs with $\mu$ can be computed,
\begin{align}
\dlog{f_i(x_i; \mu)}=\frac{\as}{\pi}\int_{x_i}^1\frac{\df z}{z}P_{ij}(z)f_j\left(\frac{x_i}{z};\mu\right)\,,
\end{align}
where the repeated index $j$ is summed over and $P_{ij}$ are the Altarelli-Parisi splitting functions. Near hadronic threshold, the splitting functions simplify and can be written as
\begin{align}
\frac{\as}{\pi}P_{ij}(x)=\left[2\ts\Gamma_{\cusp}\ts \vec{T}_i^2\ts\frac{1}{(1-x)_{+}}+\gamma_i\ts\delta(1-x)\right]\delta_{ij}\,.
\end{align}

\subsection{Soft Function}
\label{subsec:NLOsoft}

In general, the soft function depends on the null component of the soft momentum inside each jet as well as the timelike component $k^0_{\out}$ (defined in \sec{factorization} as $k_\out^0 \equiv p_I \cdot k_\out/\abs{p_I}$). At order $\as$, the soft function can be written as a sum of functions that depend only on one momentum variable, with trivial dependence on the others and is given by
\begin{equation}
S(\kout,\{n_i, k_i^+\})=S_{\out}(\kout)\prod_{i\in\jetset}\delta(k_i^+)+\delta(\kout)\sum_{i\in\jetset}S_{\ins}(k_i^+)\prod_{\substack{j\in\jetset\\j\neq i}}\delta(k_j^+)\,,
\end{equation}
where the sum over $i\in\jetset$ is over all $i$ corresponding to outgoing jets and does not include the incoming partons (and we remind the reader that the dependence here on  $k_i$ is only over final state jets but the $n_i$ run over all initial and final partons). 
The timelike component of the soft function, $S_{\out}$, receives contributions from soft gluons that are not inside any of the outgoing jets. We can find this by calculating the contribution of soft gluons going anywhere and then subtracting the contribution from gluons that enter one of the jets. This can be written in the hadronic center-of-mass frame as 
\begin{align}
S_{\out}(\kout)=&-\sum_{\pair{i}{j}}\vec{T}_i\mcdot\vec{T}_j\ts 2g^2\mu^{2\epsilon}\int\frac{\df^dk}{(2\pi)^d}\frac{n_i\mcdot n_j}{(n_i\mcdot k)(n_j\mcdot k)}\new
&\qquad \times2\pi\delta(k^2)\delta\Big(\frac{p_I\mcdot k}{\abs{p_I}}-\kout \Big)\bigg[1-\sum_{k\in\jetset}\hat \Theta^R_{{\rm soft},k}\bigg]\,,
\end{align}
where $\hat \Theta^R_k$ is the restriction that the gluon is in jet $k$, defined by a jet algorithm of size $R$ as in \sec{algorithm}. This is most easily calculated in the partonic center-of-mass frame. Denoting the directions and energies of the collinear partons in this frame as $\tilde{n}_i$ and $\tilde{\omega}_i$ respectively, we have that
\begin{align}
S_{\out}(\kout)=&-\sum_{\pair{i}{j}}\vec{T}_i\mcdot\vec{T}_j\ts 2g^2\mu^{2\epsilon}\int\frac{\df^dk}{(2\pi)^d}\frac{\tilde{n}_i\mcdot\tilde{n}_j}{(\tilde{n}_i\mcdot k)(\tilde{n}_j\mcdot k)}2\pi\delta(k^2)\delta(k^0-\kout)
\bigg[1-\sum_{k\in\jetset}\hat{\Theta}^R_{{\rm soft},k}\bigg]\,,
\end{align}
where we have used the fact that, for an $\eta-\phi$ algorithm, the jet algorithm restrictions are frame invariant. 

The null components of the soft function, $S_{\ins}(k_i^+)$, are defined in the hadronic center-of-mass frame as
\begin{align}
S_{\ins}(k_k^+)=-\sum_{\pair{i}{j}}\vec{T}_i\mcdot\vec{T}_j\ts 2g^2\mu^{2\epsilon}\int\frac{\df^dk}{(2\pi)^d}\frac{n_i\mcdot n_j}{(n_i\mcdot k)(n_j\mcdot k)}2\pi\delta(k^2)\delta(n_k\mcdot k-k_k^+)\hat \Theta^R_{{\rm soft},k}\,.
\end{align}
In the partonic center-of-mass frame, this can be written as
\begin{align}
S_{\ins}(k_k^+)=-\sum_{\pair{i}{j}}\vec{T}_i\mcdot\vec{T}_j\ts 2g^2\mu^{2\epsilon}\int\frac{\df^dk}{(2\pi)^d}\frac{\tilde{n}_i\mcdot\tilde{n}_j}{(\tilde{n}_i\mcdot k)(\tilde{n}_j\mcdot k)}2\pi\delta(k^2)\frac{\omega_k}{\tilde{\omega}_k}\delta\left(k^+-\frac{\omega_k}{\tilde{\omega}_k}k_k^+\right)\hat \Theta^R_{{\rm soft},k}\,.
\end{align}

The calculation of $S_{\ins}$ and $S_{\out}$ in the partonic center-of-mass frame can be related to the calculation of the soft function in \cite{Ellis:2010rw}. While \cite{Ellis:2010rw} uses a polar angle measure, we will show that, even though $S_{\ins}$ and $S_{\out}$ separately depend on the parameter $R$, the anomalous dimension is $R$ independent. This implies that all algorithms with the same singularity structure as the polar angle algorithm used in \cite{Ellis:2010rw} have the same anomalous dimension for this observable. Specifically, the anomalous dimension calculated in the partonic center-of-mass frame should be the same for both a polar angle measure and an $\eta-\phi$ measure. Since the $\eta-\phi$ measure is boost invariant, the $R$ independence of the anomalous dimension must be true in all frames\footnote{We have verified this explicitly for small $R$ by making the replacement $R\to R/\text{cosh}\ts\eta$, which relates the polar angle and $\eta-\phi$ measures in the small $R$ limit.}. The relations of $S_{\ins}$ and $S_{\out}$ to the soft function in \cite{Ellis:2010rw} are given by
 
\begin{equation}
\label{Soutdef}
S_{\out}(\kout)=2\sum_{\pair{i}{j}}\frac{\df}{\df\Lambda}\left[S^{\incl}_{ij}+\sum_{k\in\jetset}S^k_{ij}\right]_{\Lambda=\kout}\,,
\end{equation}
and
\begin{align}
\label{Sindef}
S_{\ins}(k_k^+)=2\sum_{\pair{i}{j}}\frac{1}{\tilde{\omega}_k}S_{ij}^{\meas}(\tau_0^k)\,,
\end{align}
where $S^{\incl}_{ij}$, $S^k_{ij}$, and $S_{ij}^\meas$ are defined and computed in \cite{Ellis:2010rw}. In calculating $S_{ij}^\meas$, we use the definitions $\tau_0^k= k_k^+/\tilde{\omega}_k$ and $\delta_R=\delta\left(\tau_0^k-k^+/\omega_k\right)$, where $\tau_a$ and $\delta_R$ are originally defined in \cite{Ellis:2010rw}. 

The anomalous dimension of this soft function is defined as
\begin{align}
\dlog{S(k^0_{\rm out} ,\{n_i, k_i^+\}; \mu)} =&\ts \!\prod_{i\in\jetset}\! \int_0^{k_i^+} \!\!\!\!\df {k'}_{\!\! i}^+ \int_0^{k^0_{\rm out}} \!\!\!\!\df {k'}^0_{\!\!\rm out} \, \gamma_S(k^0_{\rm out}-{k'}^0_{\!\!\rm out} ,\{n_i,k_i^+ - {k'}_i^+\}; \mu)\new
&\times S({k'}^0_{\!\!\rm out} ,\{n_i,{k'}_i^+\}; \mu)
\,.\end{align}
Using the results of \cite{Ellis:2010rw}, together with Eqs.~(\ref{Soutdef}) and~(\ref{Sindef}), the result for the anomalous dimension can be written as
\begin{align}
\label{eq:gamma_S_res}
\gamma_S(\kout,\{n_i,k_i^+\}; \mu)=\sum_{i\in\jetset}\gamma_{S_i}(k_i^+; \mu)\prod_{\substack{j\in\jetset\\j\neq i}}\delta(k_j^+)\ts\delta(\kout)+\gamma_{S_{\out}}(\kout; \mu)\prod_{i\in\jetset}\delta(k_i^+)
\,.\end{align}
with
\begin{align}
\gamma_{S_i}(k_i^+; \mu) &= 2\ts\Gamma_{\cusp}\vec{T}_i^2\ts\frac{\omega_i}{\mu\ts\tilde{\omega}_i}\left(\frac{\mu\ts\tilde{\omega}_i}{k_i^+\omega_i}\right)_{\plus} \\
\gamma_{S_{\out}}(\kout; \mu) &= \Gamma_{\cusp}\sum_{\pair{i}{j}}\vec{T}_i\mcdot \vec{T}_j\Biggl(2\ts\ln\frac{\tilde{n}_i\mcdot\tilde{n}_j}{2}\Biggr)\delta(\kout) -4\ts\Gamma_{\cusp}\ts(\vec{T}_{1}^2+\vec{T}_{2}^2)\frac{1}{\mu}\left(\frac{\mu}{2\kout}\right)_{\plus}
\,.
\end{align}
\section{Consistency of Factorization to ${\cal O}(\alpha_s)$}
\label{sec:consistency}
Consistency is a nontrivial check of our factorization theorem. The factorized cross section should be independent of the factorization scale $\mu$ in the threshold limit and thus renormalization group invariant. Starting from the generic $N$-jet cross section, \eq{Njet-fact-result}, ignoring multiplicative factors that do not affect the derivative, and using the shorthand notation
\begin{equation}
\dlog{\sigma} \equiv   \dlog{} \frac{\df\sigma}{\prod_{i}\df^2p_{J_i}^\T\df\eta_{J_i}\df \Phi_q}
\,,\end{equation}
we have that
\begin{align}
\label{mudep_xsection}
\dlog{\sigma}\propto &\int_\tau^1\df z\int_{-\frac{1}{2}\ln \frac{z}{\tau}}^{\frac{1}{2}\ln\frac{z}{\tau}}\df Y H(\{n_i,\omega_i\}; \mu)
f_1(x_1; \mu)f_2(x_2; \mu) \nn\\
&\times \prod_{i\in\jetset}\int\df p_i^+J_i(p_i^+; \mu) \prod_{i\in\jetset}\int\df k_i^+\int\df k^0_{\out}\softfn\new
&\times\biggl(\dlog{\ts\ln H(\{n_i,\omega_i\}; \mu)}+\dlog{\ts\ln f_1(x_1; \mu)}
+\dlog{\ts\ln f_2(x_2; \mu)} \nn\\
& \qquad \qquad +\sum_{i\in\jetset}\dlog{\ts\ln J_i(p_i^+; \mu)}+\dlog{\ts\ln S(k^0_{\rm out},\{n_i, k_i^+\}; \mu)}\biggr)\new
&\times\delta\left[1-z-\frac{2}{\sqrt{\hat{s}}}\Bigl(\cosh Y\sum_{i\in\jetset}(p_i^++k_i^+)+k_{\out}^0\Bigr)\right]
\,.\end{align}
There are several simplifications we can make to check the independence of $\mu$. First, $\mu$ only enters perturbative expressions, and whether or not the cross section depends on $\mu$ is independent of nonperturbative physics. This allows us to use the perturbative definition of the parton distribution functions. Second, given that the $\mu$ dependence of each of the factorization ingredients starts at order $\alpha_s$, we can use the tree level expressions for the hard, jet and soft functions, as well as for the PDFs,
\begin{align}
f_i(x; \mu)&=\delta(1-x)\\
H(\{n_i,\omega_i\}; \mu) &= 1\\
J_i(p_i^+; \mu)&=\delta(p_i^+)\\
\softfn&=\delta(k_{\out}^0)\prod_{i\in\jetset}\delta(k_i^+)
\,.\end{align}
Using this and working to lowest order in $\as$, we can simplify Eq.~(\ref{mudep_xsection}) to get
\begin{align}
\dlog{\sigma} \propto&\ts\frac{\as}{\pi}P_{11}(\tau)+\frac{\as}{\pi}P_{22}(\tau)+\frac{\sqrt{\hat{s}}}{2\cosh Y}\sum_{i\in\jetset}\gamma_{J_i}\left(\frac{\sqrt{\hat{s}}}{2\cosh Y}(1-\tau); \mu\right)\new
&+\frac{\sqrt{\hat{s}}}{2\cosh Y}\sum_{i\in\jetset}\gamma_{S_i}\left(\frac{\sqrt{\hat{s}}}{2\cosh Y}(1-\tau); \mu\right)+\gamma_H(\mu)\ts\delta(1-\tau)\new
&+\frac{\sqrt{\hat{s}}}{2}\gamma_{S_{\out}}\left(\frac{\sqrt{\hat{s}}}{2}(1-\tau); \mu\right).
\end{align}
After plugging in \eqss{gamma_H_res}{gamma_J_res}{gamma_S_res}, rescaling the plus functions using the identity
\begin{align}
\left(\frac{1}{ax}\right)_{\plus}=\frac{\ln a}{a}\delta(x)+\frac{1}{a}\left(\frac{1}{x}\right)_{\plus}
\,,\end{align}
and combining the various terms, we find
\begin{align}
\label{eq:mudep_xsec_res2}
\dlog{\sigma}\propto&\, \Gamma_{\cusp} \left[\sum_{\ijpair}\vec{T}_i\mcdot\vec{T}_j\left(2\ts\ln\frac{\tilde{n}_i\mcdot\tilde{n}_j}{n_i\mcdot n_j}\right)+\vec{T}_1^2\ts\ln\frac{\omega_1^2}{\hat{s}}+\ts\vec{T}_2^2\ts\ln\frac{\omega_2^2}{\hat{s}}+\sum_{i\in\jetset}\vec{T}_i^2\left(2\ts\ln\frac{\omega_i}{\tilde{\omega}_i}\right)\right]
\,.\end{align}
After making the simplification 
\begin{equation}
\sum_{\ijpair}\vec{T}_i\mcdot\vec{T}_j\left(2\ts\ln\frac{\tilde{n}_i\mcdot\tilde{n}_j}{n_i\mcdot n_j}\right)=\sum_{\substack{i,j\\ i\neq j}}\vec{T}_i\mcdot\vec{T}_j\left(\ln\frac{\omega_i}{\tilde{\omega}_i}+\ln\frac{\omega_j}{\tilde{\omega}_j}\right)=\sum_{i\in\partset}\vec{T}_i^2\left(-2\ts\ln\frac{\omega_i}{\tilde{\omega}_i}\right)\,,
\end{equation}
where we have used that $\tilde{p}_i\mcdot\tilde{p}_j=p_i\mcdot p_j$ and $\sum_i\vec{T}_i=0$, \eq{mudep_xsec_res2} gives
\begin{align}
\dlog{\sigma}=0
\,.\end{align}
This result confirms that our factorization theorem is consistent at hadronic threshold and justifies using the renormalization group to resum logarithms of $1-\tau$.
\section{Conclusions and Outlook}
\label{sec:conclusion}

We have derived a factorization theorem for the production of $N$ jets, together with any number of non-strongly interacting particles, such as electroweak gauge bosons. This factorization theorem allows us to write the physical cross section in terms of a convolution of parton distribution functions, a hard function, and jet functions for each observed jet and a soft function describing among other things the color recombination between the initial and final state partons. Both the jet and the soft functions depend on the precise form of the jet algorithm chosen. 

The main new ingredient in this factorization theorem is a soft function that depends on a timelike component of the soft momentum outside of the observed jets, and the lightlike component of the soft momentum in a given jet. This function is directly related to the soft function first proposed and calculated for the case of jet production in $e^+ e^-$ collisions in~\cite{Ellis:2010rw,Ellis:2009wj}. This soft function allows us to interpolate between the soft function arising for final states without observed jets (which depends only on a timelike component of the soft momentum) and the soft function for completely inclusive jet production (which depends only on the lightlike component of the soft momentum). 

We have derived the UV divergent parts of all ingredients of the factorization theorem to ${\cal O}(\alpha_s)$. These were then used to show that the combination of all ingredients of the factorization theorem is independent of the arbitrary factorization scale $\mu$, and therefore the derived results satisfy the nontrivial requirement of consistency. While consistency was shown in this work only in the true hadronic endpoint, we have kept the kinematics general enough (in particular allowing for a nonzero overall boost) to allow for a generalization of our results to  the case where the steepness of the parton luminosities force events to be close to the partonic threshold. This result, which is by far more interesting phenomenologically, will be the subject of future work~\cite{threshold2}. 

Our results can be used to explicitly resum threshold logarithms to NLL accuracy (in the log-counting convention of \cite{Becher:2007ty}) for any process in hadron collisions with any number of jets and non-strongly interacting particles in the final state. The technology of going from the anomalous dimensions we present here to explicit resummed distributions is well-known. Beyond one jet, in addition to the standard resummation methods, we need the matrices $\vecT_i \cdot \vecT_j$, but these have been computed for many processes  (see, \eg, \cite{Chiu:2009mg,Kidonakis:1998bk,Kidonakis:1998nf,Dokshitzer:2005ek,Sjodahl:2008fz,Sjodahl:2009wx}), including all $2\to2$ and $2\to3$ partonic channels. The only other ingredient needed to obtain a NLL distribution is the Born matrix element. 

In addition, if our results are extended to include two-loop results of the anomalous dimensions together with the full algorithm-dependent one-loop finite parts, NNLL results can be obtained for all processes for which the virtual NLO corrections are known. Together with recent advances in calculations of NLO cross sections (\eg, $W^+W^-j$ \cite{Campbell:2003hd,Dittmaier:2007th} and $Wjjj$ \cite{Berger:2009ep}), this would have a significant impact on the precision frontier of predictions at the LHC.

\begin{acknowledgments}
This work was supported by the Director, Office of Science, and Offices of High Energy and Nuclear Physics of the U.S. 
Department of Energy under the Contract No. DE-AC02-05CH11231. A.H. also acknowledges support from an LHC Theory Initiative Graduate Fellowship, NSF Grant No. PHY-0705682.
\end{acknowledgments}


\bibliography{bibliography}

\end{document}